\documentclass[aps,prl,twocolumn,showpacs,superscriptaddress,groupedaddress,nofootinbib]{revtex4-1}

\usepackage{amsmath}
\usepackage{amsfonts}
\usepackage{dsfont}
\usepackage{enumerate}
\usepackage{color}
\usepackage{graphicx}
\usepackage{bm}
\usepackage{amssymb}
\usepackage[colorlinks=true,citecolor=blue,urlcolor=blue]{hyperref}

\begin{document}
\title{Constraints on Scattering of keV--TeV Dark Matter with Protons in the Early Universe}
\author{Vera Gluscevic}
\affiliation{School of Natural Sciences, Institute for Advanced Study, Einstein Drive, Princeton, NJ 08540, USA}
\author{Kimberly K. Boddy}
\affiliation{
Department of Physics \& Astronomy, Johns Hopkins University, Baltimore, MD 21218, USA\\
Department of Physics \& Astronomy, University of Hawaii, Honolulu, HI 96822, USA
}

\begin{abstract}
We present the first cosmological constraint on dark matter scattering with protons in the early Universe for the entire range of dark matter masses between 1\,keV and 1\,TeV.
This constraint is derived from the \textit{Planck} measurements of the cosmic microwave background (CMB) temperature and polarization anisotropy, and the CMB lensing anisotropy.
It improves upon previous CMB constraints by many orders of magnitude, where limits are available, and closes the gap in coverage for low-mass dark matter candidates.
We focus on two canonical interaction scenarios: spin-independent and spin-dependent scattering with no velocity dependence.
Our results exclude (with 95\% confidence) spin-independent interactions with cross sections greater than $5.3$$\times$$10^{-27}$\,cm$^2$ for 1\,keV, $3.0$$\times$$10^{-26}$\,cm$^2$ for 1\,MeV, $1.7$$\times$$10^{-25}$\,cm$^2$ for 1\,GeV, and $1.6$$\times$$10^{-23}$\,cm$^2$ for 1\,TeV dark matter mass.
Finally, we discuss the implications of this study for dark matter physics and future observations.
\end{abstract}

\pacs{}
\maketitle

\textit{Introduction.}
One of the primary ways to investigate the fundamental nature of dark matter (DM) is to search for evidence of its nongravitational interactions with the standard model of particle physics.
None of the experimental or observational searches for DM interactions have yet made a confirmed detection.
As a result, large portions of DM parameter space are excluded and new proposals are coming on the scene to broaden the search strategy and examine unexplored DM scenarios~\cite{Alexander:2016aln,Battaglieri:2017aum}.

A substantial effort to detect and characterize the properties of DM rests on the hypothesis that DM may be a weak-scale thermal relic particle (WIMP) with a mass in the range of tens of GeV to a few TeV~\cite{Jungman:1995df}.
Virtually all traditional direct-detection searches are constructed and optimized to search for WIMPs from the local Galactic halo through their scattering on nuclei in underground targets~\cite{2013arXiv1310.8327C}.
They have exquisite sensitivity: the tightest constraints to date on spin-independent interactions exclude cross sections greater than $\sim$$8\times 10^{-47}$\,cm$^2$ for masses around 30\,GeV~\cite{2017PhRvL.119r1301A,2017arXiv171003572F}, while the next-generation experiments promise to push this bound further in the near future~\cite{2013arXiv1310.8327C}.
These searches, however, are looking under the lamp post.
Specifically, current nuclear-recoil--based measurements are effectively blind or background dominated below DM masses of about a GeV~\cite{2013arXiv1310.8327C}.
Additionally, the extensive shielding inherent in underground experiments puts a ``ceiling'' on the interaction strength, above which the majority of particles would be stopped before reaching the detector~\cite{2005PhRvD..72h3502Z,Emken:2017qmp,2017JCAP...12..004S}.
Technological improvements~\cite{Angloher:2015ewa,Agnese:2017jvy} and analyses of electronic recoils are able to expand direct-detection sensitivities to somewhat lower DM masses~\cite{2012PhRvL.109b1301E,2017PhRvD..96d3017E}; however, the latter are only applicable to DM interactions with electrons, not protons. Entirely new experimental strategies are thus required to truly open up sub-GeV DM to broad, in-depth exploration that parallels dedicated WIMP searches~\cite{Alexander:2016aln,Battaglieri:2017aum}.

In addition to direct detection, there is a range of studies that constrain low-energy DM--baryon interactions in the local Universe, using results from balloon-borne experiments~\cite{2007PhRvD..76d2007E}, Galactic structure~\cite{2001sddm.symp..263W}, observations of galaxy clusters~\cite{2008MNRAS.384..814H,2002ApJ...580L..17N}, cosmic rays~\cite{2002PhRvD..65l3503C,2017APS..APR.J5008C}, and other astrophysical observations~\cite{1990PhRvD..41.3594S,2005A&A...438..419B,2007PhRvD..76d3523M,2017arXiv171204901K}.
These studies explore various parts of the DM parameter space, but few focus specifically on sub-GeV particles.

Given the current null results, new DM models (\textit{e.g.,} hidden-sector DM~\cite{Feng:2008ya}, asymmetric DM~\cite{Kaplan:2009ag}, freeze-in DM~\cite{Hall:2009bx}, SIMPs/ELDERs~\cite{Hochberg:2014dra,Hochberg:2014kqa,Kuflik:2015isi}) have recently received much attention in theoretical and experimental communities.
Many of these models comfortably accommodate DM particles with masses in the keV--GeV range.
In this study, we produce the strongest cosmological constraint to date on DM interactions covering the entire DM mass range between 1\,keV and 1\,TeV, using \textit{Planck} measurements of the cosmic microwave background (CMB) temperature and polarization anisotropy, and the CMB lensing anisotropy~\cite{2016A&A...594A...1P,2016A&A...594A..11P}.\footnote{Lyman-$\alpha$ forest limits on warm DM exclude masses below a few keV~\cite{Viel:2013fqw}, and we thus focus only on masses above this limit.}

\begin{figure}[t]
\includegraphics[height=.35\textwidth]{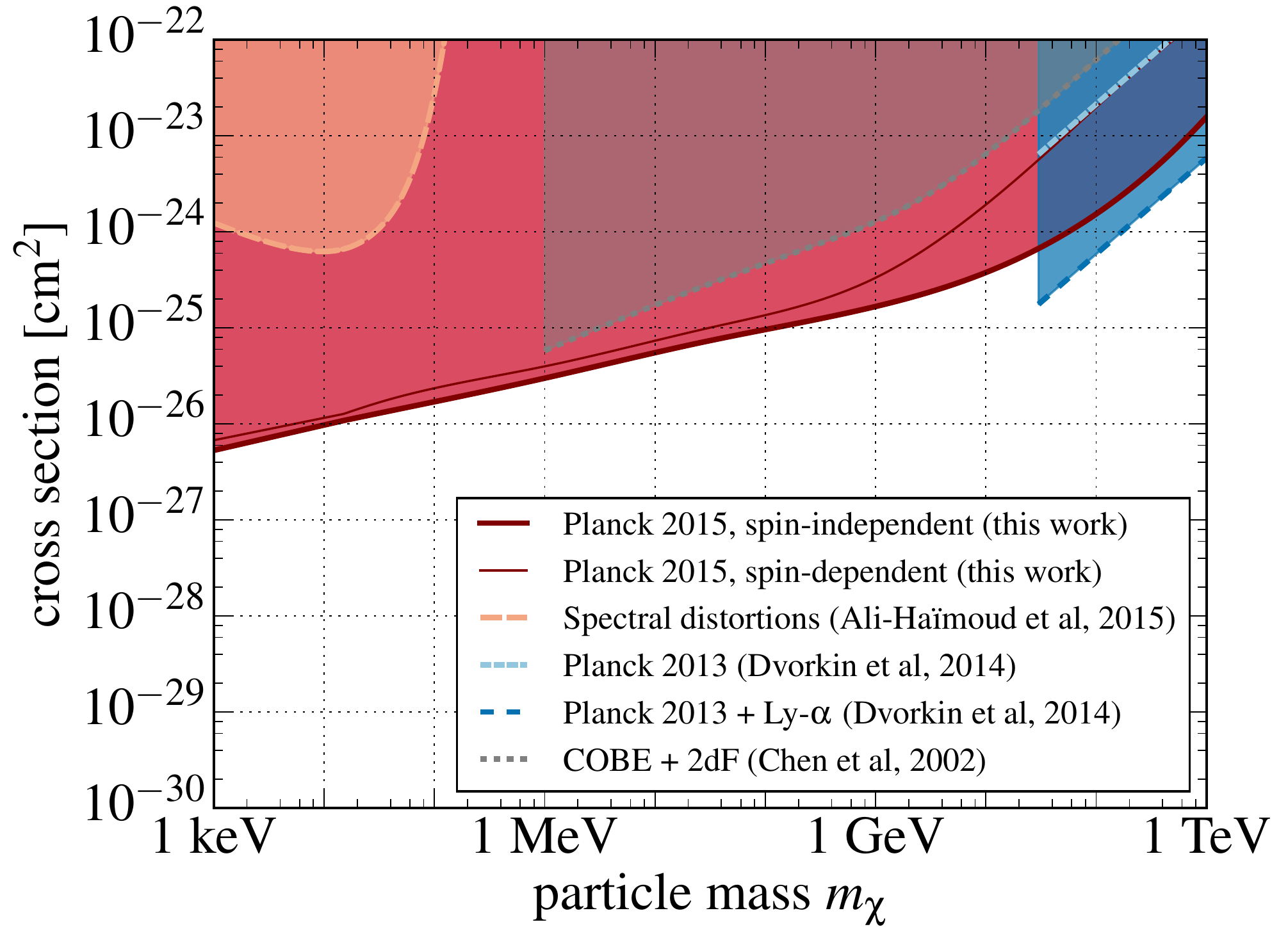}
\caption{Constraints on the DM--proton scattering cross section, as derived from various cosmological measurements; shaded regions are excluded with 95\% confidence.
The exclusion curves that partially span this mass range are from previous state-of-the-art results, while the red curves that span the entire mass range represent the constraints derived in this study for spin-independent and spin-dependent scattering.
  \label{fig:constraints}}
\end{figure}

The very same interactions sought locally by direct detection and other experiments also take place in the early Universe (in the first $\sim$400 000 years after the Big Bang) and can be tested with cosmological observations.
If baryons scatter with DM particles in the primordial plasma prior to recombination, the heat transferred to the DM fluid can cool the photons, producing spectral distortions in the CMB; this effect was previously used to constrain DM masses below a few hundred keV from the null detection of distortions in FIRAS data~\cite{Ali-Haimoud:2015pwa} (see also Figure~\ref{fig:constraints}).
Furthermore, due to a drag force between the DM and photon--baryon fluids, small-scale matter fluctuations are suppressed, altering the shape of the CMB power spectra and of the matter power spectrum.
This effect too was explored in previous studies~\cite{Chen:2002yh,Sigurdson:2004zp,Dvorkin:2013cea} and was most recently used to place constraints for DM much heavier than a GeV~\cite{Dvorkin:2013cea} (see again Figure~\ref{fig:constraints}).

We expand upon this previous work in several important ways.
First, we probe DM with masses down to 1\,keV, thereby closing the gap in mass coverage of previous cosmological studies.
Additionally, only scattering with free protons in the early Universe was previously considered for the leading cosmological constraints on heavy DM~\cite{Dvorkin:2013cea}; we also account for DM scattering with helium nuclei, which significantly improves constraints in that mass regime.
Finally, we use the latest \textit{Planck} 2015 data release~\cite{2016A&A...594A...1P}, and for the first time include CMB polarization and lensing measurements to search for evidence of DM--proton interactions.
With these improvements in our analysis, the limits we obtain are a factor of $\sim$13 stronger than the best previous CMB limits of Ref.~\cite{Dvorkin:2013cea} for heavy DM.\footnote{When we make the same simplifying assumptions as Ref.~\cite{Dvorkin:2013cea}, we restore consistency with their results.}
In addition, our constraints are several orders of magnitude stronger than those of Refs.~\cite{Chen:2002yh,Ali-Haimoud:2015pwa} for lower DM masses.\footnote{The constraint of Ref.~\cite{Chen:2002yh} is not dominated by the CMB measurements, but rather by a reconstruction of the linear matter power spectrum from the 2dF galaxy survey~\cite{2001Natur.410..169P}, which may strongly depend on the choice of galaxy bias model.}

\textit{Dark matter--proton scattering.}
We concentrate on two DM--proton interaction scenarios: spin-independent and spin-dependent elastic scattering, with no dependence on relative particle velocity.
These simple interactions are the most widely considered and easily arise at leading order from high-energy theories (the literature on this subject is vast, and we refer the reader to an early review for reference~\cite{Jungman:1995df}).
In a companion paper~\cite{Boddy:2017inprep}, we expand this study to constrain DM--proton interactions in the broader context of nonrelativistic effective field theory~\cite{Fan:2010gt,Fitzpatrick:2012ix,Anand:2013yka} and address a wide range of momentum- and velocity-dependent interactions.

In order to compute CMB power spectra in the presence of the interactions, we modify the code \texttt{CLASS}~\cite{Blas:2011rf} to solve the following Boltzmann equations (in synchronous gauge)~\cite{Ma:1995ey}
\begin{equation}
\begin{gathered}
  \dot{\delta}_\chi = -\theta_\chi - \frac{\dot{h}}{2}, \ \dot{\delta}_b = -\theta_b    - \frac{\dot{h}}{2}\\
  \dot{\theta}_\chi = -\frac{\dot{a}}{a}\theta_\chi + c_\chi^2 k^2 \delta_\chi + R_\chi (\theta_b - \theta_\chi)\\
  \dot{\theta}_b    = -\frac{\dot{a}}{a}\theta_b   + c_b^2 k^2 \delta_b + R_\gamma (\theta_\gamma - \theta_b)+ \frac{\rho_\chi}{\rho_b} R_\chi (\theta_\chi - \theta_b)
\end{gathered}
\label{eq:boltzmann}
\end{equation}
for the evolution of DM and baryon density fluctuations, ${\delta}_\chi$ and ${\delta}_b$, and velocity divergences, ${\theta}_\chi$ and ${\theta}_b$, respectively.
In the above expressions, $k$ is the wave number of a given Fourier mode; $a$ is the scale factor; $h$ is the trace of the scalar metric perturbation~\cite{Ma:1995ey}; ${c}_b$ and $c_\chi$ are the speeds of sound in the two fluids~\cite{Ma:1995ey}; and $\rho_b$ and $\rho_\chi$ are their respective energy densities.
The overdot notation represents a derivative with respect to conformal time.
The subscript $\gamma$ pertains to photons, where $R_\gamma$ represents the usual Compton scattering term~\cite{Ma:1995ey}.

The terms proportional to $R_\chi$ encapsulate the new interaction physics;
$R_\chi$ is the coefficient for the rate of momentum exchange between the DM and baryon fluids, found by averaging the momentum-transfer cross section over the velocity distributions of particles in the early Universe~\cite{Sigurdson:2004zp,Dvorkin:2013cea}.
Previous work considered DM scattering with only free protons~\cite{Dvorkin:2013cea}; here, we include scattering with protons inside helium nuclei, and thus need a more general expression for $R_\chi$ to account for the nuclear structure of helium.

We start by summarizing the results for scattering with free protons.
In this case, both the spin-independent (SI) and spin-dependent (SD) cross sections are the same as the corresponding momentum-transfer cross sections,
\begin{align}
  \sigma_p^{(\text{SI})} &= \frac{\mu_{\chi p}^2}{m_v^4 \pi}
  \left[c_p^{(\text{SI})}\right]^2 \nonumber\\
  \sigma_p^{(\text{SD})} &= \frac{\mu_{\chi p}^2}{m_v^4 \pi}
  \frac{S_\chi(S_\chi + 1)}{4} \left[c_p^{(\text{SD})}\right]^2 \ ,
\end{align}
where $S_\chi$=$1/2$ is the spin of the DM, $m_\chi$ is the mass of the DM particle, and $\mu_{\chi p}$ is the reduced mass of the DM--proton system.
The coupling coefficients $c_p^{(\text{SI})}$ and $c_p^{(\text{SD})}$ set the strength of the spin-independent and spin-dependent interactions, respectively.
We insert the weak-scale mass $m_v$$\approx$246\,GeV, as an overall normalization.\footnote{The choice of the normalization scale does not impact our constraints on the cross sections.}

Moving on to helium, we first note that it has zero spin and thus cannot have spin-dependent interactions.
For the spin-independent interaction, there is no inherent velocity dependence; however, the nuclear form factor is a function of the momentum transferred in the scattering process\footnote{The momentum transfer is given by $q^2$=$2\mu_{\chi\text{He}}^2 v^2 (1-\cos\theta)$, where $\theta$ is the scattering angle in the center-of-mass frame, $v$ is the relative velocity between the DM and helium particles, and $\mu_{\chi\text{He}}$ is the reduced mass of the DM--helium system.}~\cite{Catena:2015uha}.
Thus, the associated momentum-transfer cross section has a velocity-dependent part multiplying the following numerical factor
\begin{equation}
  \sigma_\text{He}^{(\text{SI})} = 4\frac{\mu_{\chi\text{He}}^2}{m_v^4 \pi}
  \left[c_p^{(\text{SI})}\right]^2 \ ,
\end{equation}
which depends on the strength of the interaction, quantified by $c_p^{(\text{SI})}$, and $\mu_{\chi\text{He}}$ is the reduced mass of the DM--helium system.
When we average the full momentum-transfer cross sections (multiplied by the relative particle velocity) over the velocity distributions for DM and baryons, we obtain\footnote{Since our constraints imply thermal decoupling of DM and baryons at early times $z$$>$$10^{4}$, we are able to neglect the relative bulk velocity between the DM and baryon fluids to arrive at this expression~\cite{Dvorkin:2013cea,Munoz:2015bca}.}
\begin{align}
  R_{\chi p}^{(\text{SI/SD})}
  &= \mathcal{N}_0 a\rho_b (1-Y_\text{He})
  \frac{\sigma_p^{(\text{SI/SD})}}{m_\chi+m_p}
  \left(\frac{T_b}{m_p}+\frac{T_\chi}{m_\chi}\right)^{\frac{1}{2}} \nonumber\\
  R_{\chi\text{He}}^{(\text{SI})}
  &= \mathcal{N}_0 a\rho_b Y_\text{He}
  \frac{\sigma_\text{He}^{(\text{SI})}}{m_\chi+m_\text{He}}
  \left(\frac{T_b}{m_\text{He}}+\frac{T_\chi}{m_\chi}\right)^{\frac{1}{2}}
  \nonumber\\ & \hphantom{=} \times \left[1+(2\mu_\text{He} a_\text{He})^2
    \left(\frac{T_b}{m_\text{He}}+\frac{T_\chi}{m_\chi} \right)\right]^{-2},
  \label{eq:Rchi}
\end{align}
where $\mathcal{N}_0$$\equiv$$2^{\frac{7}{2}}/3\sqrt{\pi}$, $Y_\text{He}$ is the helium mass fraction, and $T_b$ and $T_\chi$ are the temperatures of the baryon and DM fluids.
In the nuclear shell model, the length parameter for helium is $a_\text{He}$$\approx$$1.5$~fm~\cite{Fitzpatrick:2012ix}.
For spin-independent scattering, the total rate coefficient is $R_\chi^{(\text{SI})} = R_{\chi p}^{(\text{SI})} + R_{\chi\text{He}}^{(\text{SI})}$; for spin-dependent scattering, the total rate coefficient is $R_\chi^{(\text{SD})} = R_{\chi p}^{(\text{SD})}$.
Note that the velocity dependence of the cross section in the case of helium translates to the additional temperature-dependent term in the last line of the above expressions.

Since we are interested in light DM, we cannot neglect terms with $T_\chi$ in the above equations (as was done in Ref.~\cite{Dvorkin:2013cea} for heavy DM).
We thus track the DM temperature evolution given by\footnote{At early times, when the interactions affect the evolution of density modes accessible to cosmological observables, baryons are in thermal contact with photons, and the backreaction on the baryon temperature is a subdominant effect; we thus ignore it.}~\cite{Sigurdson:2004zp,Dvorkin:2013cea}
\begin{equation}
\begin{gathered}
\dot{T}_\chi = -2\frac{\dot{a}}{a}T_\chi+ 2 R^\prime_\chi (T_b - T_\chi) \ .
\end{gathered}
\label{eq:Tchi}
\end{equation}
The heat-exchange coefficients control when the DM and baryon fluids thermally decouple, and they are given by
\begin{equation}
\begin{gathered}
R_\chi^{\prime (\text{SI})} \equiv (\mu_{\chi p}/m_p) R_{\chi p}^{(\text{SI})}+(\mu_{\chi\text{He}}/m_\text{He})R_{\chi\text{He}}^{(\text{SI})} \ , \\
R_\chi^{\prime (\text{SD})} \equiv (\mu_{\chi p}/m_p) R_{\chi p}^{(\text{SD})} \ .
\end{gathered}
\label{eq:Rchiprime}
\end{equation}

\textit{Data analysis and results.}
\begin{figure}[t]
\includegraphics[height=.34\textwidth]{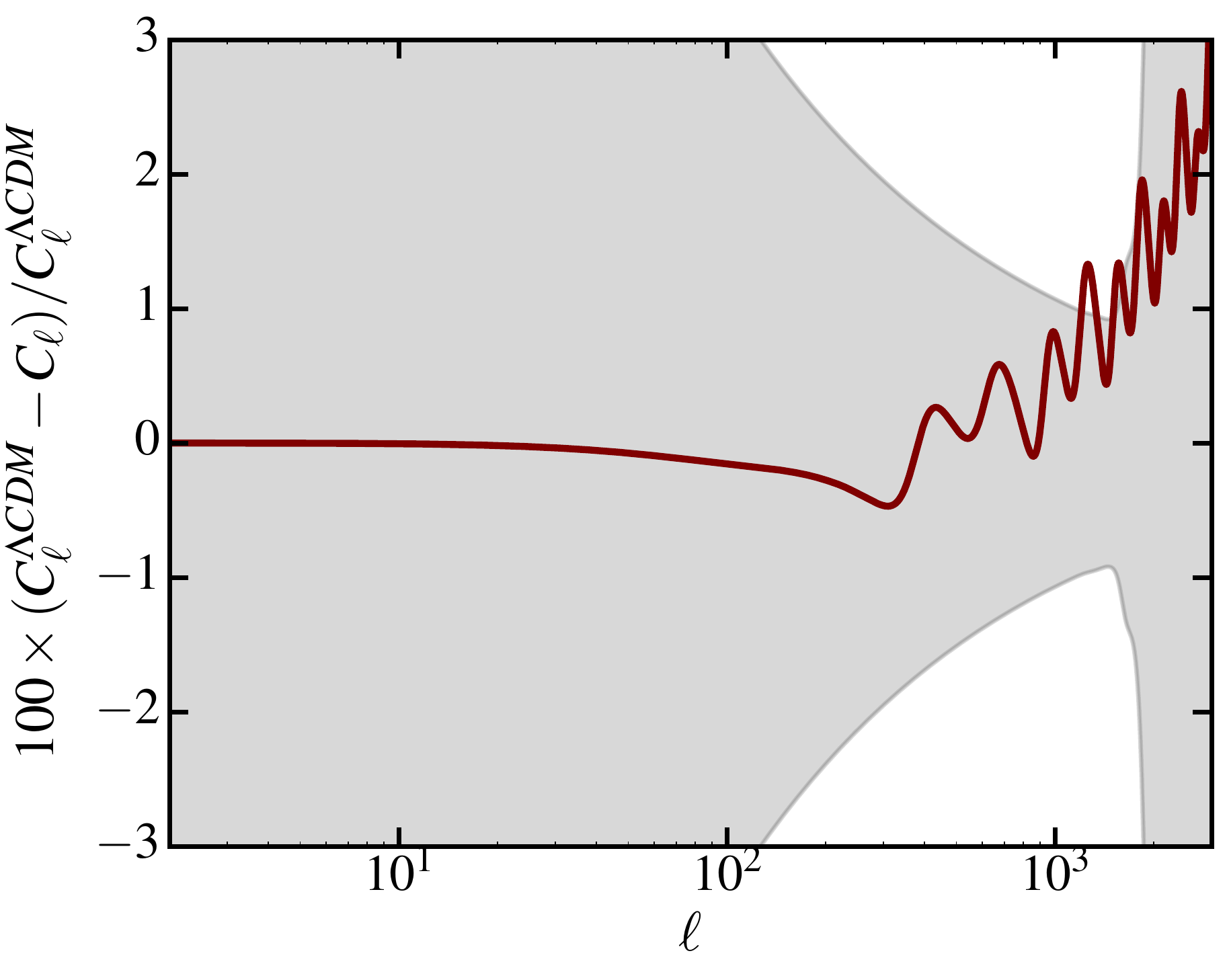}
\caption{Percent difference in the CMB temperature power spectrum between the $\Lambda$CDM model and a model with spin-independent DM--proton scattering, where the interaction strength is set to its $95\%$ confidence-level upper limit (while all other cosmological parameters are kept at their best-fit \textit{Planck} 2015 values~\cite{2016A&A...594A..13P}).
  The size of \textit{Planck} $2\sigma$ error bar (binned with a bin size $\Delta\ell$=$50$) is roughly represented by the shaded region, for reference.
  Note that most of \textit{Planck}'s constraining power comes from the smallest well-measured angular scale at $\ell$$\sim$1200; roughly speaking, this implies that our reported constraint is most sensitive to DM scattering at a redshift corresponding to the horizon entry of the smallest measurable mode that has experienced scattering with baryons for the longest time.
  \label{fig:residuals}}
\end{figure}
We use the CMB power spectra and likelihoods from the \textit{Planck} 2015 data release, as available through the \texttt{clik/plik} distribution~\cite{2016A&A...594A..11P,2016A&A...594A...1P}.
We analyze temperature, lensing, and low-$\ell$ polarization to jointly constrain the six standard $\Lambda$CDM parameters: the Hubble parameter $h$, baryon density $\Omega_bh^2$, DM density $\Omega_\chi h^2$, reionization optical depth $\tau$, the amplitude of the scalar perturbations $A_s$, and the scalar spectral index $n_s$.
We also include the coupling coefficient $c_p^{\text{SI/SD}}$ as an additional free parameter (with a wide flat prior probability distribution).
We use the code \texttt{MontePython}~\cite{Audren:2012wb} with the \texttt{PyMultinest}~\cite{2014A&A...564A.125B} implementation of nested likelihood sampling~\cite{Feroz:2007kg,Feroz:2008xx,Feroz:2013hea}.\footnote{For the case of no DM--proton interactions (vanishing coupling coefficients), we recover $\Lambda$CDM parameter values and constraints consistent with \textit{Planck} published results~\cite{2016A&A...594A...1P} (to within $0.14\sigma$).}
We repeat the fitting procedure for a range of 8 fixed DM mass values between 1\,keV and 1\,TeV for spin-independent and for spin-dependent interactions.\footnote{We choose to fix the mass, rather than to sample it as a free parameter, purely for computational reasons. The results are not affected by this choice; an equivalent approach would be to vary the mass as a free parameter and report the 95\% confidence-level contours of the marginalized posterior in mass-cross section parameter space as an exclusion curve.}

We find no evidence for DM--proton scattering in the data, and thus derive 95$\%$ confidence-level upper limits on $c_p^{\text{SI}}$ and $c_p^{\text{SD}}$ as a function of DM mass.
We then convert these results into upper limits on the corresponding interaction cross sections; the resulting exclusion curves are shown and compared to previous results\footnote{The results of Ref.~\cite{Dvorkin:2013cea} are only valid for $m_\chi$$\gg$$m_p$. The slope of their constraint starts to deviate noticeably from our exact calculation at $\sim$50\,GeV.} in Figure~\ref{fig:constraints}.
For the spin-independent interaction, we exclude cross sections greater than $5.3$$\times$$10^{-27}$\,cm$^2$ for 1\,keV, $3.0$$\times$$10^{-26}$\,cm$^2$ for 1\,MeV, $1.7$$\times$$10^{-25}$\,cm$^2$ for 1\,GeV, and $1.6$$\times$$10^{-23}$\,cm$^2$ for 1\,TeV DM particle mass.
To illustrate the effect of scattering, Figure~\ref{fig:residuals} shows the percent difference in the CMB temperature power spectrum between the $\Lambda$CDM model and a model with spin-independent DM--proton scattering.

Most of the constraining power in this analysis comes from the temperature measurements.
The scattering signal appears as a similar suppression of power at high multipoles in the case of lensing and polarization, but since the small--scale anisotropy in these observables is not measured with high enough accuracy with \textit{Planck}, they only contribute to the limits at the level of $\sim$$30\%$.
On the other hand, while the inclusion of scattering on helium makes only a modest contribution for sub-GeV DM masses, it improves the limits by as much as a factor of 6 at high masses (in Figure~\ref{fig:constraints}, compare the spin-independent limit and spin-dependent limit; helium contributes only to the former).
This is a consequence of the mass dependence of the momentum-transfer rate between DM and baryons. With helium included, the maximal momentum-transfer rate occurs at a higher mass (by a factor of a few, as compared to the proton-only case).
Given the rapid loss of sensitivity with increasing mass (see Figure~\ref{fig:constraints}), this shift implies modest improvements in constraining power at masses around a GeV, but substantial improvements in the high-mass regime.

Finally, the scaling of the cross-section constraint with DM mass depends on two quantities that enter all relevant evolution equations: $R_\chi$ and $R_\chi^{\prime}$.
For heavy DM, both rates scale as $\sim$$\sigma_p/m_\chi$, as does the resulting exclusion curve shown in Figure~\ref{fig:constraints}; thus, our result can be directly extended to higher masses by appropriately scaling our reported limit at 1\,TeV.
In the low-mass limit, the mass scaling of the rates is different [see Eqs.~\eqref{eq:Rchi} and \eqref{eq:Rchiprime}], and the slope of the exclusion curve is a nontrivial combination of the two effects.

\textit{Conclusions.}
We analyze \textit{Planck} measurements of temperature, polarization, and lensing anisotropy to perform the first cosmological search for dark matter--proton scattering in the early Universe in the full range of dark matter masses between 1\,keV and 1\,TeV.
We find no evidence of such interactions and thus report an upper bound on the corresponding cross sections, shown in Figure~\ref{fig:constraints}.
This analysis improves upon previous leading CMB limits by one or more orders of magnitude, for masses where they were available.

We directly constrain cross sections for dark matter scattering with protons---the same quantities probed by direct detection and other experiments that operate at low energies, but extend to a regime in parameter space that is inaccessible to current underground experiments.
Additionally, upper limits coming from all experimental probes seeking to detect dark matter in the Galactic halo are sensitive to the assumptions about the astrophysical properties of dark matter particles (their local velocity distribution and energy density, in particular).
The limits we report \textit{directly} address cosmological dark matter in the early Universe and thus sidestep these important caveats of the local low-energy probes.
Therefore, our result provides highly complementary information on dark matter interaction physics, and paves the road for a broad approach to the dark matter problem.

The effect of dark matter interactions is progressively more prominent at smaller angular scales (see Figure~\ref{fig:residuals}), making it a prime target of investigation for a number of existing and upcoming low-noise, high-resolution, ground-based CMB experiments, such as the Atacama Cosmology Telescope (ACT)~\cite{2017JCAP...06..031L}, the South Pole Telescope (SPT)~\cite{2017arXiv170709353H}, the Simons Observatory\footnote{\url{https://simonsobservatory.org}}, and the CMB Stage-4 experiment~\cite{2016arXiv161002743A}.
We expect a substantial improvement in the sensitivity of our analysis with data from ground-based CMB measurements in the near future.

\textit{Acknowledgments.}~VG gratefully acknowledges the support of the Eric Schmidt fellowship at the Institute for Advanced Study.
We thank Yacine Ali-Ha\"{i}moud, John Beacom, Jo Dunkley, Ely Kovetz, and Samuel McDermott for comments on the manuscript, and Marc Kamionkowski, Jason Kumar, Zack Li, Joel Meyers, Vivian Poulin, and David Spergel for useful discussions. We especially thank Tracy Slatyer and the anonymous referees for useful feedback on the original manuscript.

\bibliography{physics-refs}
\end{document}